# PageRank algorithm for Directed Hypergraph

Loc Tran[1], Tho Quan[2], and An Mai[3]

[1]John von Neumann Institute, VNU-HCM, Ho Chi Minh City, Vietnam (E-mail: tran0398@umn.edu)
[2]Ho Chi Minh City University of Technology, VNU-HCM, Ho Chi Minh City, Vietnam (E-mail: qttho@hcmut.edu.vn)
[3]John von Neumann Institute, VNU-HCM, Ho Chi Minh City, Vietnam (E-mail: an.mai@jvn.edu.vn)

**Abstract**

During the last two decades, we easily see that the World Wide Web's link structure is modeled as the directed graph. In this paper, we will model the World Wide Web's link structure as the directed hypergraph. Moreover, we will develop the PageRank algorithm for this directed hypergraph. Due to the lack of the World Wide Web directed hypergraph datasets, we will apply the PageRank algorithm to the metabolic network which is the directed hypergraph itself. The experiments show that our novel PageRank algorithm is successfully applied to this metabolic network.

## 1 Introduction

With the quick growth of the World Wide Web during the last two decades, information retrieval introduces growing theoretical and practical challenges. With the huge amount of information inflowing the World Wide Web every second, it becomes more difficult and more difficult to retrieve information from the Web. This explains why the existence of a search engine is as important as the existence of the web itself. Since the appearance of the web, there has been a fundamental talk in the web research community to develop the rapid, effective, and precise search engines.

This paper will be chiefly discussing about the most common search engine nowadays which is Google. The mathematical theory behind the Google search engine is the PageRank algorithm, which was presented by Sergey Brin and Lawrence Page [1]. In 1998, Brin and Page were PhD students at Stanford University, USA. Then they took a leave of absence from their Ph.D. to concentrate on developing their Google model. Their original paper describing the PageRank algorithm is used nowadays by Google to create the rankings of the web pages of the World Wide Web.

A search engine contains three important modules: a crawler, and an indexer, and a query engine [2]. The crawler gathers and stores data from the web. Data is stored in the indexer which mines information from the data gathered from the crawler. The query engine responds to the queries from customers. The PageRank algorithm (i.e. one of the ranking algorithms), part of the query engine, ranks the web pages in the order of their "importance" to the query. The ranking is attained by the contribution of a score to each web page of the World Wide Web.

PageRank is a ranking algorithm of web pages of the World Wide Web. The PageRank algorithm exploits the link structure of the web. The World Wide Web's link structure forms a directed graph where the web pages are the nodes of the directed graph and the links are the directed edges of the directed graph. The web page (i.e. the node of the directed graph) is considered "important" if it is pointed to by other "important" web pages.

During the last two decades, we easily recognized that the World Wide Web's link structure was modeled as the directed graph. Moreover, the PageRank algorithm was developed for this directed graph only. In general, this model, the directed graph, is not the best and the generalized model for the World Wide Web's link structure. In this paper, we will model the World Wide Web's link structure as the directed hypergraph [3, 4]. This work, to the best of our knowledge, has not been investigated up to now. However, due to the lack of the World Wide Web directed hypergraph datasets, we will exploit the metabolic network dataset that is available from [5]. This metabolic network can easily be represented as the directed graph or the directed hypergraph. We will show clearly how to construct the directed graph and the directed hypergraph from this metabolic network. Then our next task is to develop the PageRank algorithm for the directed hypergraph. This is the novel work. Moreover, we will define the un-normalized and the symmetric normalized directed hypergraph Laplacian in this paper. In the future, if the World Wide Web's directed hypergraph datasets are available, then the directed hypergraph Laplacian based semi-supervised learning will be developed in order to solve the spam detection problem. We can easily see that the applications of the directed hypergraph are huge.

We will organize the paper as follows: Section 2 will introduce the preliminary notations and definitions used in this paper. Section 3 will introduce the PageRank algorithm for the directed hypergraph. Section 4 will introduce the definitions of the un-normalized and symmetric normalized directed hypergraph Laplacian and their applications. Section 5 will show the experimental results. Section 6 will conclude this paper and discuss the future direction of researches.

## 2 Preliminary notations and definitions

Given the directed hypergraph $H = (V, E)$ where $V$ is the set of vertices and $E$ is the set of hyper-arcs. Each hyper-arc $e \in E$ is written as $e = (e^{Tail}, e^{Head})$. The vertices of $e$ are denoted by $\boldsymbol{e} = e^{Tail} \cup e^{Head}$. $e^{Tail}$ is called the tail of the hyper-arc $e$ and $e^{Head}$ is called the head of the hyper-arc $e$. Please note that $e^{Tail} \neq \emptyset, e^{Head} \neq \emptyset, e^{Tail} \cap e^{Head} = \emptyset$.

The directed hypergraph $H = (V, E)$ can be represented by two incidence matrices $H^{Tail}$ and $H^{Head}$.

These two incidence matrices $H^{Tail}$ and $H^{Head}$ can be defined as follows

$$h^{Tail}(v, e^{Tail}) = \begin{cases} 1 \; if \; v \in e^{Tail} \\ 0 \; otherwise \end{cases}$$

$$h^{Head}(v, e^{Head}) = \begin{cases} 1 \; if \; v \in e^{Head} \\ 0 \; otherwise \end{cases}$$

Let $w(e)$ be the weight of the hyper-arc $e$. Let $W$ be the diagonal matrix containing the weights of hyper-arcs in its diagonal entries.

From the above definitions, we can define the tail and head degrees of the vertex v and the tail and head degrees of the hyper-arc e as follows

$$d^{Tail}(v) = \sum_{e \in E} w(e) h^{Tail}(v, e^{Tail})$$

$$d^{Head}(v) = \sum_{e \in E} w(e) h^{Head}(v, e^{Head})$$

$$d^{Tail}(e) = \sum_{v \in V} h^{Tail}(v, e^{Tail})$$

$$d^{Head}(e) = \sum_{v \in V} h^{Head}(v, e^{Head})$$

Let $D_v^{Tail}$, $D_v^{Head}$, $D_e^{Tail}$, and $D_e^{Head}$ be four diagonal matrices containing the tail and head degrees of vertices and the tail and head degrees of hyper-arcs in their diagonal entries respectively. Please note that $D_v^{Head}$ and $D_v^{Tail}$ are the $R^{|V|*|V|}$ matrices and $D_e^{Head}$ and $D_e^{Tail}$ are the $R^{|E|*|E|}$ matrices.

**3 PageRank algorithm for directed hypergraph**

From [4], we know that the transition probability of the random walk on directed hypergraph can be defined as follows

$$p(u, v) = \sum_{e \in E} w(e) \frac{h^{Tail}(u, e^{Tail})}{d^{Tail}(u)} \frac{h^{Head}(v, e^{Head})}{d^{Head}(e)}$$

From the above definition, the transition probability matrix $P$ of the random walk on the directed hypergraph can be defined as follows

$$P = {D_v^{Tail}}^{-1} H^{Tail} W {D_e^{Head}}^{-1} {H^{Head}}^{T}$$

Next, we need to prove that the row sum of $P$ is 1.

We have that

$$\sum_{v\in V} p(u,v) = \sum_{v\in V}\sum_{e\in E} w(e)\frac{h^{Tail}(u,e^{Tail})}{d^{Tail}(u)}\frac{h^{Head}(v,e^{Head})}{d^{Head}(e)}$$
$$= \sum_{e\in E} w(e)\frac{h^{Tail}(u,e^{Tail})}{d^{Tail}(u)}\frac{1}{d^{Head}(e)}\sum_{v\in V} h^{Head}(v,e^{Head})$$
$$= \sum_{e\in E} w(e)\frac{h^{Tail}(u,e^{Tail})}{d^{Tail}(u)}$$
$$= \frac{1}{d^{Tail}(u)}\sum_{e\in E} w(e)h^{Tail}(u,e^{Tail})$$
$$= 1$$

So, we easily see that $p(u,v) \geq 0, \forall u,v$ and $\sum_{v\in V} p(u,v) = 1, \forall v$. Then we can conclude that $P$ is the stochastic matrix. Following the work from [1], the PageRank vector $\pi$ of the directed hypergraph is the left dominant eigenvector of the transition probability matrix $P$ of the random walk on the directed hypergraph. In the other words, the PageRank vector $\pi$ of the directed hypergraph is the solution of the following equation

$$\pi^T = \pi^T P$$

Moreover, we know that the above equation can easily be solved by the Power method.

## 4 Directed hypergraph Laplacian

In this section, we give two novel definitions of the directed hypergraph Laplacian which are un-normalized directed hypergraph Laplacian and symmetric normalized directed hypergraph Laplacian.

Let $S$ be the diagonal matrix containing all elements of PageRank vector $\pi$ of the directed hypergraph in its diagonal entries.

The un-normalized directed hypergraph Laplacian can be defined as follows

$$L = S - \frac{S{D_v^{Tail}}^{-1}H^{Tail}W{D_e^{Head}}^{-1}{H^{Head}}^T + ({D_v^{Tail}}^{-1}H^{Tail}W{D_e^{Head}}^{-1}{H^{Head}}^T)^T S}{2}$$

The symmetric normalized directed hypergraph Laplacian can be defined as follows

$$L_{sym} = I - \frac{S^{\frac{1}{2}}{D_v^{Tail}}^{-1}H^{Tail}W{D_e^{Head}}^{-1}{H^{Head}}^T S^{-\frac{1}{2}} + S^{-\frac{1}{2}}({D_v^{Tail}}^{-1}H^{Tail}W{D_e^{Head}}^{-1}{H^{Head}}^T)^T S^{\frac{1}{2}}}{2}$$

From these two definitions, we can develop the directed hypergraph Laplacian Eigenmaps algorithms, the spectral directed hypergraph clustering algorithms, and the directed hypergraph Laplacian based semi-supervised learning algorithms. These are our future works.

## 5 Experiments and Results

Datasets

Due to the lack of the World Wide Web directed hypergraph datasets, we use the metabolic network dataset that is available from [5]. This metabolic network is itself the directed hypergraph. Thus, we don't need to transform it to the directed hypergraph. This metabolic network contains 72 metabolites (i.e. nodes of the directed hypergraph) and 95 reactions (i.e. hyper-arcs of the directed hypergraph). However, in the experiment, we just use 50 metabolites that are both in the head AND in the tail of some hyper-arcs. In the other words, we can avoid the case that the tail degree OR the head degree of the metabolite is zero. Moreover, we just use 75 reactions that the tail degree AND the head degree of the hyper-arc are not zero. In the other words, we set that $e^{Tail} \neq \emptyset$ AND $e^{Head} \neq \emptyset$.

Finally, we have the metabolic network (i.e. the directed hypergraph dataset) that has 50 metabolites (i.e. nodes of the dataset) and 75 reactions (i.e. hyper-arcs of the dataset).

## Experiments

In the experiment, we initially construct $H^{Tail}$ and $H^{Head}$ matrices. Then we can construct $D_v^{Tail}$, $D_v^{Head}$, $D_e^{Tail}$, and $D_e^{Head}$ matrices. Finally, we can compute the transition probability matrix $P$ of the random walk on the directed hypergraph.

In order to obtain the PageRank vector of the directed hypergraph, we compute the left dominant eigenvector of the matrix $P$. The following table 1 and table 2 show the 10 highest rank values of 10 metabolites in the directed hypergraph and the names of these 10 metabolites that have highest ranks.

| Rank order | Rank values |
|---|---|
| 1 | 0.6366 |
| 2 | 0.2640 |
| 3 | 0.2321 |
| 4 | 0.2180 |
| 5 | 0.2087 |
| 6 | 0.2039 |
| 7 | 0.2006 |
| 8 | 0.1941 |
| 9 | 0.1798 |
| 10 | 0.1701 |

Table 1: The rank values of 10 highest ranks

| Rank order | Names of metabolites with highest ranks |
|---|---|
| 1 | H |
| 2 | Nicotinamide-adenine-dinucleotide-reduced |
| 3 | ADP |
| 4 | Phosphate |
| 5 | ATP |
| 6 | Nicotinamide-adenine- |
| | dinucleotide-phosphate |
| 7 | H |
| 8 | Pyruvate |
| 9 | Nicotinamide-adenine-dinucleotide |
| 10 | Coenzyme-A |

Table 2: The names of 10 metabolites that have highest ranks

**6 Conclusions**

In this paper, we develop the PageRank algorithm for the directed hypergraph. We successfully apply it to the metabolic network which is the directed hypergraph itself. This is the novel work not only in web mining field but also in bio-informatics field. In the future, we will develop the directed hypergraph Laplacian based semi-supervised learning in order to solve the spam detection problem since two directed hypergraph Laplacian was defined in this paper.

Moreover, in the future, we will also try to develop the directed hypergraph p-Laplacian semi-supervised learning method. This method is worth investigated because of its hard nature and its close connection to PDE on directed hypergraph field.

**Acknowledgement**

This research is funded by Vietnam National University Ho Chi Minh City (VNU-HCM) under grant number B2018-20-07.